\documentclass[a4paper,12pt]{article}
\usepackage[latin1]{inputenc}

\usepackage{cite}
\usepackage{graphicx}
\usepackage{psfrag}
\usepackage{amsmath}
\usepackage{geometry}
\title{Cooperative Spectrum Sensing Using\\
Random Matrix Theory}

\author{
\begin{footnotesize}
\centering
\begin{tabular}{cc}
	Leonardo S. Cardoso and Merouane Debbah and Pascal Bianchi & Jamal Najim\\
	SUPELEC & ENST\\
	Gif-sur-Yvette, France & Paris, France\\
	\{leonardo.cardoso, merouane.debbah and pascal.bianchi\}@supelec.fr & jamal.najim@enst.fr\\
\end{tabular}
\end{footnotesize}
}

\begin{document}

\maketitle

\begin{abstract}
In this paper, using tools from asymptotic random matrix theory, a
new cooperative scheme for frequency band sensing is introduced for
both AWGN and fading channels. Unlike previous works in the field,
the new scheme does not require the knowledge of the noise
statistics or its variance and is related to the behavior of
the largest and smallest eigenvalue of random matrices. Remarkably,
simulations show that the asymptotic claims hold even for a small
number of observations (which makes it convenient for time-varying
topologies), outperforming classical energy detection techniques.
\end{abstract}

\section{Introduction}

It has already become a common understanding that current
mobile communication systems do not make full use of the available
spectrum, either due to sparse user access or to the system's
inherent deficiencies, as shown by a report from the Federal
Communications Commission (FCC) Spectrum Policy Task
Force~\cite{rpt-fcc-2002}. It is envisioned that future systems will
be able to opportunistically exploit those spectrum 'left-overs', by
means of knowledge of the environment and cognition capability, in
order to adapt their radio parameters accordingly. Such a technology
has been proposed by Joseph Mitola in 2000 and is called cognitive
radio \cite{phd-mitola-2000}. Due to the fact that recent advances
on micro-electronics and computer systems are pointing to a -not so
far- era when such radios will be feasible, it is of utmost
importance to develop good performing sensing techniques.

In its simplest form, spectrum sensing means looking for a signal in
the presence of noise for a given frequency band (it could also
encompass being able to classify the signal). This problem has been
extensively studied before, but it has regained attention now as part
of the cognitive radio research efforts. There are several classical techniques
for this purpose, such as the energy detector (ED)\cite{art-urkowitz-1967,
art-kostylev-2002, art-digham-2003}, the matched filter \cite{art-sahai-2003}
and the cyclostationary feature detection \cite{inproc-cabric-2004,
art-fehske-, art-tang-}. These techniques have their strengths and
weaknesses and are well suited for very specific applications.

Nevertheless, the problem of spectrum sensing as seen from a cognitive
radio perspective, has very stringent requirements and limitations,
such as,

\begin{itemize}

    \item no prior knowledge of the signal structure (statistics, noise
    variance value, etc...);

    \item the detection of signals in the shortest time possible;

    \item ability to detect reliably even over heavily faded environments;

\end{itemize}

The works by Cabric et al.~\cite{inproc-cabric-2004}, Akyildiz et
al.~\cite{art-akyildiz-2006} and Haykin~\cite{art-haykin-2005}
provide a summary of these classical techniques from the cognitive
network point of view. It is clear from these works, that none can
fully cope with  all the requirements of the cognitive radio
networks.

In simple AWGN (Additive White Gaussian Noise) channels, most
classical approaches perform very well. However, in the case of fast
fading, these techniques are not able to provide satisfactory
solutions, in particular to the hidden node problem
\cite{art-fullmer-1997}. To this end, several works
\cite{inproc-mishra-2006, inproc-ganesan-2005, art-ghasemi-2005,
art-ghasemi-2007} have looked into the case in which cognitive
radios cooperate for sensing  the spectrum. These works aim at
reducing the probability of false alarm by adding extra redundancy
to the sensing process. They also aim at reducing the number of
samples collected, and thus, the estimation times by the use
parallel measuring devices. However, even though one could exploit
the spatial dimension efficiently, these works are based on the same
fundamental techniques, which require a priori knowledge of the
signal.

In this work, we introduce an alternative method for blind (in the
sense that no a priori knowledge is needed) spectrum sensing. This
method relies on the use of multiple receivers to infer on the
structure of the received signals using random matrix theory (RMT). We
show that we can estimate the spectrum occupancy reliably with a small
amount of received samples.

The remainder of this work is divided as follows. In
section~\ref{sec_prob_formulation}, we formulate the problem of
blind spectrum sensing. In section~\ref{sec_spectrum_sensing_rmt},
we introduce the proposed approach based on random matrix theory. In
section~\ref{sec_perf_anal}, we present some practical results which
confirm that the asymptotic assumptions hold even for a small amount
of samples. Then, in section~\ref{sec_results}, we show the
performance results of the proposed method. Finally, in section
\ref{sec_conclusions}, we draw the main conclusions and point out
further studies.

\section{Problem Formulation}\label{sec_prob_formulation}

The basic problem concerning spectrum sensing is the detection of a
signal within a noisy measure. This turns out to be a difficult
task, especially if the received signal power is very low due to
pathloss or fading, which in the blind spectrum sensing case is
unknown. The problem can be posed as a hypothesis test such that
\cite{art-urkowitz-1967}:

\begin{equation}\label{eq_hypothesis_test}
y(k)= \left\lbrace\begin{array}{ll}
n(k) \text{:} & H_0\\
h(k)s(k) + n(k) \text{:} & H_1\\
\end{array}
\right.,
\end{equation}

\noindent
where $y(k)$ is the received vector of samples at instant
$k$, $n(k)$ is a noise (not necessarily gaussian) of variance
$\sigma^2$, $h(k)$ is the fading component, $s(k)$ is the signal
which we want to detect, such that $E\left[\mid s(k)\mid^2
\right]\neq0$, and $H_0$ and $H_1$ are the noise-only and signal
hypothesis, respectively. We suppose that the channel $h$ stays
constant during $N$ blocks ($k=1..N$).

Classical techniques for spectrum sensing based on energy detection
compare the signal energy with a known threshold $V_T$
\cite{art-urkowitz-1967, art-kostylev-2002, art-digham-2003} derived
from the statistics of the noise and channel. The following is
considered to be the decision rule

\begin{displaymath}
\text{decision}=\left\lbrace\begin{array}{ll}
H_0,  & \text{if}\: E\left[ \mid y(k) \mid ^2\right] < V_T\\
H_1,  & \text{if}\: E\left[ \mid y(k) \mid ^2\right] \geq V_T\\
\end{array}
\right.,
\end{displaymath}

\noindent where $E[\mid y(k) \mid ^2 ]$ is the energy of the signal
and $V_T$ is usually taken as the noise variance. One drawback of
this approach is that neither the noise/channel distribution nor
$V_T$ are known a priori. In real life scenarios $V_T$ depends on
the radio characteristics and is hard to be estimated properly.
Moreover, in the case of fading and path loss, the energy of the
received signal can be of the order of the noise, making it
difficult to be detected all the more as the number of samples $N$
may be very limited. Indeed, $E\left[ \mid y(k) \mid ^2\right]$ is
estimated by

\begin{displaymath}
 \frac{1}{N} \sum_{k=1}^N \mid y(k) \mid ^2,
\end{displaymath}

\noindent
which is not a good estimator for the small sample size case.

In the following, we provide a cooperative approach for cognitive networks
to detect the signal from a primary system without the need to know
the noise variance using results from random matrix theory.

\section{Random Matrix Theory for Spectrum Sensing}\label{sec_spectrum_sensing_rmt}

\begin{figure}[!h]
 \centering
 \psfrag{primary}{\small{primary base station}}
 \psfrag{BS1}{\small{$BS_1$}}
 \psfrag{BS2}{\small{$BS_2$}}
 \psfrag{BS3}{\small{$BS_3$}}
 \psfrag{BSk}{\small{$BS_K$}}
 \includegraphics[width=0.5\columnwidth]{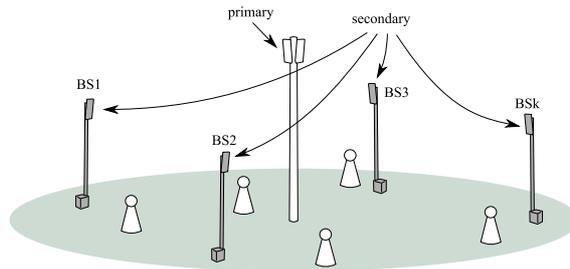}
 \caption{Considered scenario for spectrum sensing.}
 \label{fig_scenario}
\end{figure}

Consider the scenario depicted in Figure \ref{fig_scenario}, in which primary users (in white) communicate
to their dedicated (primary) base station. Secondary base stations $\{BS_1, BS_2, BS_3, ..., BS_K\}$ are cooperatively sensing the channel in order to identify a white space and exploit the medium.

Before going any further, let us assume the following:

\begin{itemize}

 \item The $K$ base stations in the secondary system share information between them. This can be performed by transmission
 over a wired high speed backbone.

 \item The base stations are analyzing the same portion of the spectrum.
\end{itemize}

Let us consider the following $K \times N$ matrix consisting of the
samples received by all the $K$ secondary base stations ($y_i(k)$ is
the sample received by base station $i$ at instant $k$):

\begin{displaymath}
{\bf Y}=\left[
\begin{array}{ccccc}
y_1(1) & y_1(2) & \cdots & y_1(N) \\
y_2(1) & y_2(2) & \cdots & y_2(N) \\
y_3(1) & y_3(2) & \cdots & y_3(N) \\
\vdots & \vdots &        & \vdots \\
y_K(1) & y_K(2) & \cdots & y_K(N) \\
\end{array}
\right].
\end{displaymath}

The goal of the random matrix theory approach is to perform a test
of independence of the signals received by the various base
stations. Indeed, in the presence of signal ($H_1$ case), all the
received samples are correlated, whereas when no signal is present
($H_0$ case), the samples are decorrelated whatever the fading
situation. Hence, in this case, for a fixe $K$ and $N \rightarrow
\infty$, the sample covariance matrix $\frac{1}{N} {\bf Y}{\bf Y}^H$
converge $\sigma^2 {\bf I}$. However, in practice, $N$ can be of the
same order of magnitude than $K$ and therefore one can not infer
directly $\frac{1}{N} {\bf Y}{\bf Y}^H$ independence of the samples.
This can be formalized using tools from random matrix theory
\cite{art-marchenko-1967}. In the case where the entries of ${\bf
Y}$ are independent (irrespectively of the specific probability
distribution, which corresponds to the case where no signal is
transmitted - $H_0$) results from asymptotic random matrix theory
\cite{art-marchenko-1967} state that:

{\bf Theorem.} Consider an $K \times N$ matrix ${\bf W}$ whose
entries are independent zero-mean complex (or real) random variables
with variance $\frac{\sigma^2}{N}$ and fourth moments of order
$O(\frac{1}{N^2})$. As  $K,N \rightarrow \infty$ with $\frac{K}{N}
\rightarrow \alpha$, the empirical distribution of $ {\bf W}{\bf
W}^H$ converges almost surely to a nonrandom limiting distribution
with density

\begin{eqnarray*}
f(x)=(1-\frac{1}{\alpha})^{+}
\delta(x)+\frac{\sqrt{(x-a)^+(b-x)^+}}{2\pi \alpha x}
\end{eqnarray*}

\noindent
where

\begin{displaymath}
    a=\sigma^2(1-\sqrt{\alpha})^2 \qquad \text{and} \qquad b=\sigma^2(1+\sqrt{\alpha})^2.
\end{displaymath}

\begin{figure}[h]
 \centering
 \psfrag{MP}{\hspace{-1mm}\v{M}P}
 \psfrag{a}{$a$}
 \psfrag{b}{$b$}
 \includegraphics[width=0.5\columnwidth]{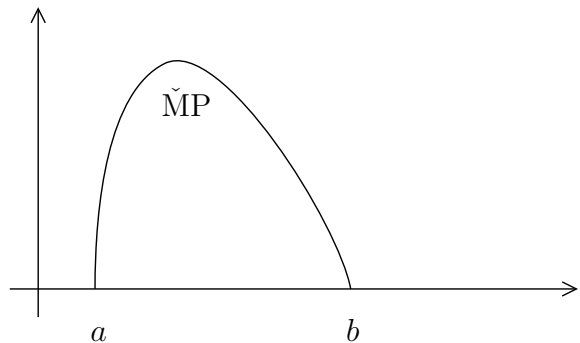}
 \caption{The Marchenko-Pastur support ($H_0$ hypothesis).}
 \label{fig_mp}
\end{figure}

Interestingly, when there is no signal, the support of the
eigenvalues of the sample covariance matrix (in Figure \ref{fig_mp},
denoted by \v{M}P) is finite, whatever the distribution of the
noise. The Marchenko-Pastur law thus serves as a theoretical
prediction under the assumption that matrix is "all noise".
Deviations from this theoretical limit in the eigenvalue
distribution should indicate non-noisy components i.e they should
suggest information about the matrix.

In the case in which a signal is present ($H_1$), {\bf Y} can be
rewritten as

\vspace{-3mm}
\begin{displaymath}
{\bf Y}=\left[
\begin{array}{ccccc}
h_1 & \sigma &  & 0 \\
\vdots &  & \ddots &  \\
h_K & 0 &  & \sigma \\
\end{array}
\right]
\left[
\begin{array}{ccc}
s(1) & \cdots & s(N) \\
z_1(1) & \cdots & z_1(N) \\
\vdots & & \vdots \\
z_K(1) & \cdots & z_K(N) \\
\end{array}
\right],
\end{displaymath}

\noindent where $s(i)$ and $z_k(i)=\sigma n_k(i)$ are respectively
the independent signal and noise with unit variance at instant $i$
and base station $k$. Let us denote by ${\bf T}$ the matrix:

\begin{displaymath}
{\bf T}=\left[
\begin{array}{ccccc}
h_1 & \sigma &  & 0 \\
\vdots &  & \ddots &  \\
h_K & 0 &  & \sigma \\
\end{array}
\right].
\end{displaymath}

${\bf T}{\bf T}^H$ has clearly one eigenvalue $\lambda_1=\sum{\vert
h_i\vert^2}+\sigma^2$ and all the rest equal to $\sigma^2$. The
behavior of the eigenvalues of $\frac{1}{N} {\bf Y}{\bf Y}^H$ is
related to the study of the eigenvalue of large sample covariance
matrices of spiked population models \cite{art-baik-2005}. Let us
define the signal to noise ratio (SNR) $\rho$ in this work as

\begin{displaymath}
 \rho = \frac{\sum{\vert h_i\vert^2}}{\sigma^2}.
\end{displaymath}

Recent works of Baik et al.\cite{art-baik-2005, art-baik-2006} have shown that, when

\begin{equation}\label{eq_baik_crit}
 \frac{K}{N} < 1 \qquad \text{and} \qquad \rho > \sqrt{\frac{K}{N}}
\end{equation}

\noindent
(which are assumptions that are clearly met when the number of
samples $N$ are sufficiently high), the maximum eigenvalue of
$\frac{1}{N} {\bf Y}{\bf Y}^H$ converges almost surely to

\begin{displaymath}
 b' = (\sum{\vert h_i\vert^2} + \sigma^2)(1+\frac{\alpha}{\rho}),
\end{displaymath}

\noindent
which is superior to $b=\sigma^2(1+\sqrt{\alpha})^2$ seen for the $H_0$ case.

\begin{figure}[h]
 \centering
 \psfrag{MP}{\hspace{-1mm}\v{M}P}
 \psfrag{a}{$a$}
 \psfrag{b}{$b$}
 \psfrag{bl}{$b'$}
 \includegraphics[width=0.5\columnwidth]{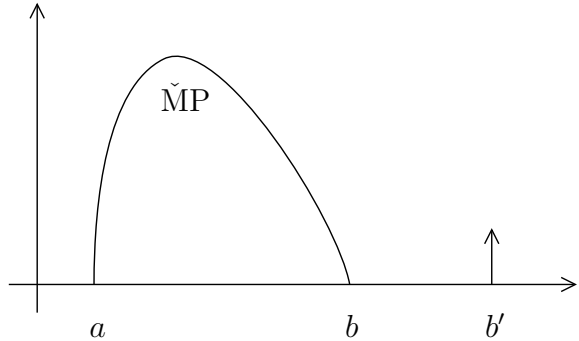}
 \caption{The Marchenko-Pastur support plus a signal component.}
 \label{fig_mp_plus_signal}
\end{figure}

Therefore, whenever the distribution of the eigenvalues of the
matrix $\frac{1}{N} {\bf Y}{\bf Y}^H$ departs from the
Marchenko-Pastur law (Figure \ref{fig_mp_plus_signal}), the detector
knows that the signal is present. Hence, one can use this
interesting feature to sense the spectrum.

Let $\lambda_i$ be the eigenvalues of $\frac{1}{N} {\bf Y}{\bf Y}^H$
and $G=[a,b]$, the cooperative sensing algorithm  works as follows:

\subsection{Noise distribution unknown, variance known}\label{sec_known}

In this case, the following criteria is used:

\begin{equation}\label{eq_decision_rmt1}
\text{decision} = \left\lbrace
\begin{array}{lc}
H_0:& \text{, if} \: \lambda_i \in G\\
H_1:&  otherwise\\
\end{array}\right.
\end{equation}

Note that refinements of this algorithm (where the probability of
false alarm is taken into account in the non-asymptotic case) can be
found in \cite{art-cardoso-2008}. The results are based on the
computation of the asymptotic largest eigenvalue distribution in the
$H_0$ and $H_1$ case.

\subsection{Both noise distribution and variance unknown}\label{sec_unknown}

Note that the ratio of the maximum and the minimum eigenvalues in
the $H_0$ hypothesis case does not depend on the noise variance. Hence, in
order to circumvent the need for the knowledge of the noise, the
following criteria is used:

\begin{equation}\label{eq_decision_rmt2}
\text{decision} = \left\lbrace
\begin{array}{lc}
H_0:& \text{, if} \: \frac{\lambda_{\text{max}}}{\lambda_{\text{min}}} \leq  \frac{(1+\sqrt{\alpha})^2}{(1-\sqrt{\alpha})^2}\\
H_1:&  otherwise\\
\end{array}\right.
\end{equation}

It should be noted that in this case, one needs to still take a
sufficiently high number of samples $N$ such that the conditions in
Eq. (\ref{eq_baik_crit}) are met. In other words, the number of
samples scales quadratically with the inverse of  the signal to
noise ratio. Note moreover that the test $H_1$ provides also a good
estimator of the SNR $\rho$. Indeed, the ratio of largest eigenvalue
($b'$) and smallest ($a$) of $\frac{1}{N} {\bf Y}{\bf Y}^H$ is
related solely to $\rho$ and $\alpha$ i.e

\begin{eqnarray*}
\frac{b'}{a}=\frac{(\rho
+1)(1+\frac{\alpha}{\rho})}{(1-\sqrt{\alpha})^2}
\end{eqnarray*}

To our knowledge, this estimator of the SNR has never been put
forward in the literature before.

\section{Performance Analysis}\label{sec_perf_anal}

The previous theoretical results have shown that one is able to
distinguish a signal from noise by the use of only a limiting ratio
of the highest to the smallest eigenvalue of the sample covariance
matrix. For finite dimensions, the operating region for such an
algorithm is still an issue and is related to the asymptotic
distribution of a scaling factor of the ratio \cite{art-cardoso-2008}.
This section provides some characterization of this region through
the analysis of the ratio between $\lambda_{max}$ and $\lambda_{min}$
of $\frac{1}{N} {\bf Y}{\bf Y}^H$ for various matrix sizes.

\begin{figure}[h]
 \centering
 \includegraphics[width=0.6\columnwidth]{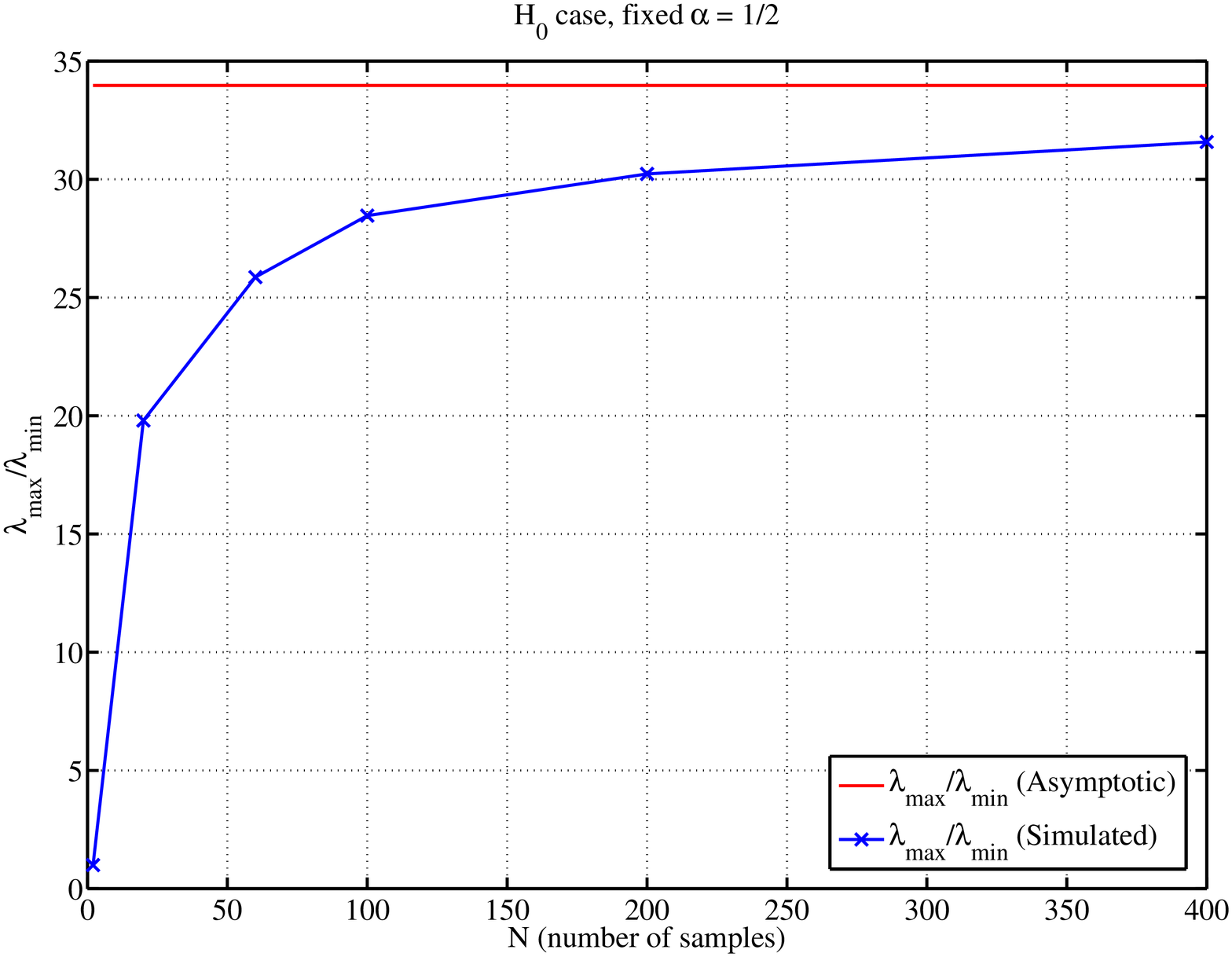}
 \caption{Behavior of $\lambda_{max}/\lambda_{min}$ for increasing $N$ (case $H_0$, $\alpha = 1/2$).}
 \label{fig_h0_asymp_0.5}
\end{figure}

\begin{figure}[h]
 \centering
 \includegraphics[width=0.6\columnwidth]{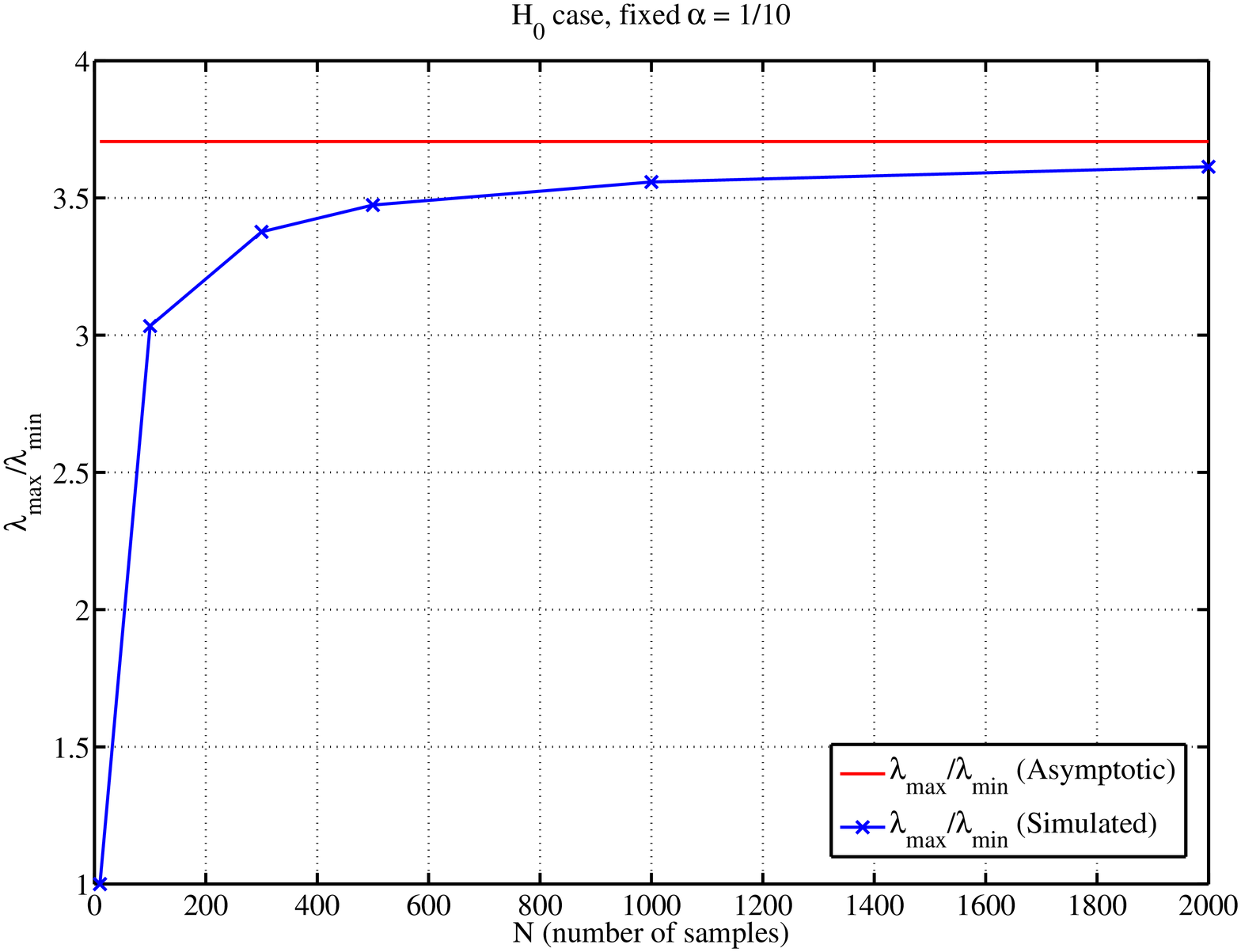}
 \caption{Behavior of $\lambda_{max}/\lambda_{min}$ for increasing $N$ (case $H_0$, $\alpha = 1/10$).}
 \label{fig_h0_asymp_0.1}
\end{figure}

Figures \ref{fig_h0_asymp_0.5} and \ref{fig_h0_asymp_0.1} present
the $\lambda_{max}/\lambda_{min}$ for various sizes of ${\bf Y}$ in
the pure noise case, with $\alpha = 1/2$ and $\alpha = 1/10$,
respectively. From the figures we see that both cases provide a good
approximation of the asymptotic ratio even with small matrix sizes.
If one takes, for example, $N=100$ ($K=50$ for $\alpha = 1/2$ and
$K=10$ for $\alpha = 1/10$), it can be seen that the simulated cases
are respectively equal to 81\% percent and 83\% of the asymptotic
limit for $\alpha = 1/2$ and $\alpha = 1/10$. As expected, for a
larger ${\bf Y}$ matrix size, the empirical ratio approaches the
asymptotic one.

\begin{figure}[h]
 \centering
 \includegraphics[width=0.6\columnwidth]{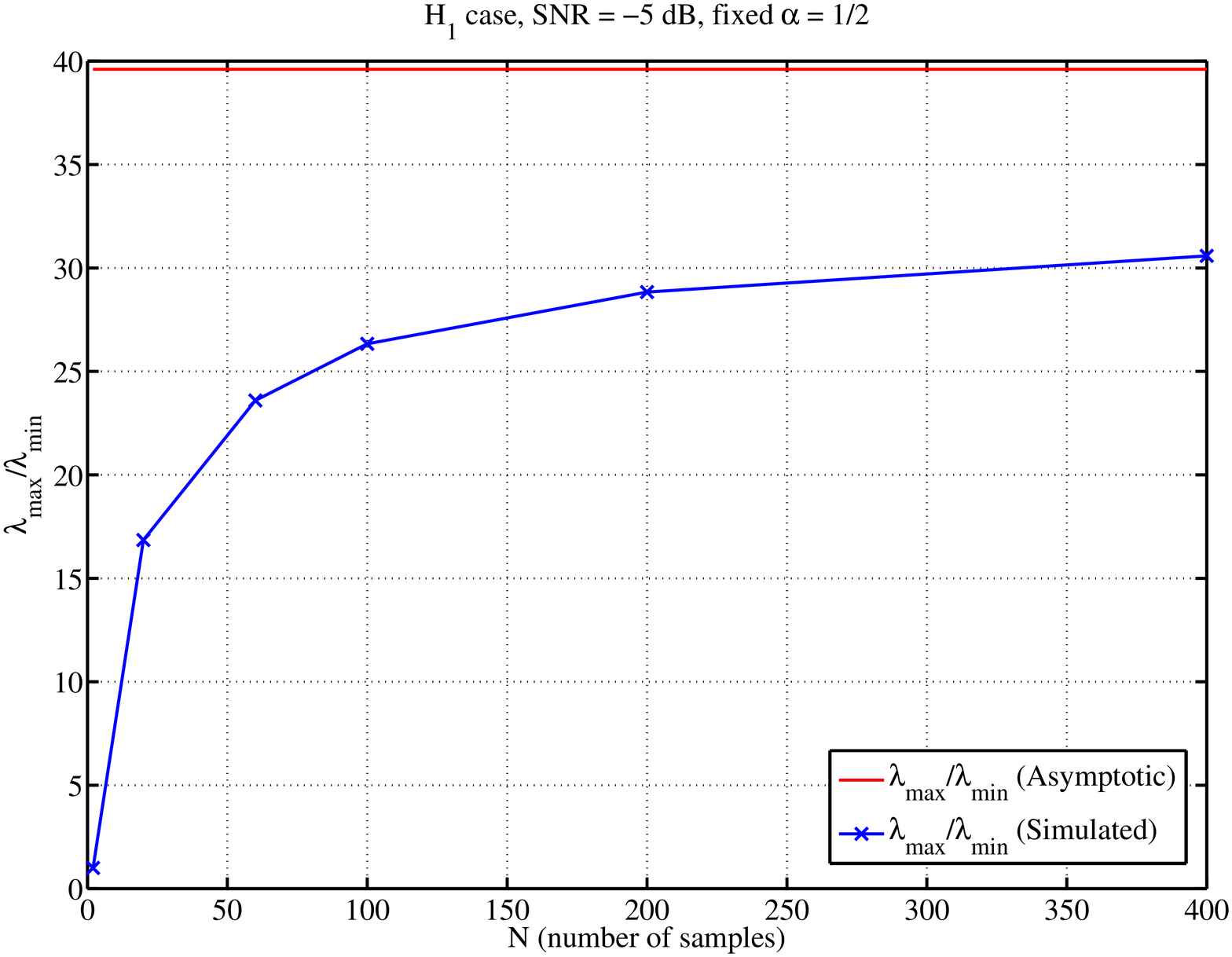}
 \caption{Behavior of $\lambda_{max}/\lambda_{min}$ for increasing $N$ (case $H_1$, $\alpha = 1/2$).}
 \label{fig_h1_asymp_0.5}
\end{figure}

\begin{figure}[h]
 \centering
 \includegraphics[width=0.6\columnwidth]{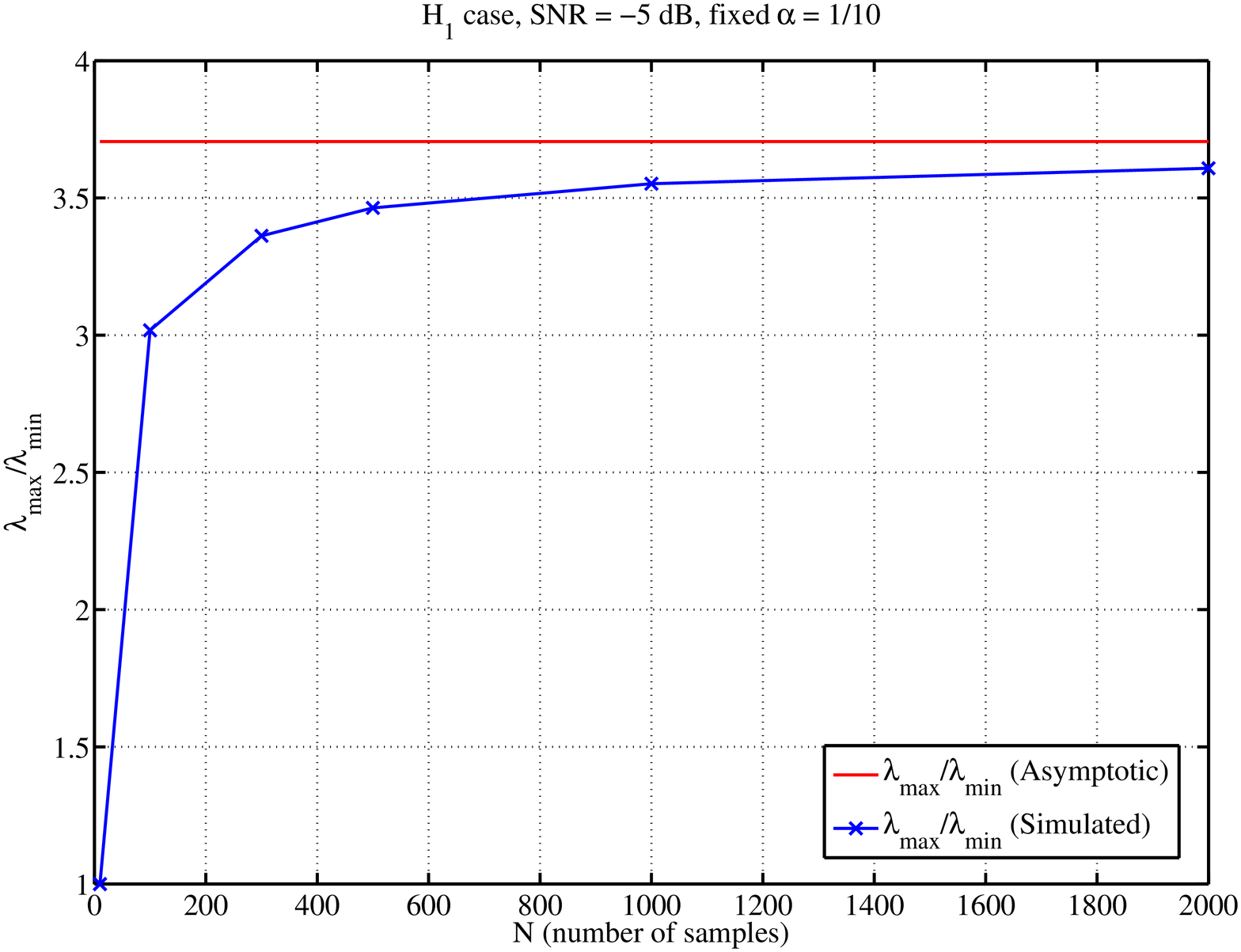}
 \caption{Behavior of $\lambda_{max}/\lambda_{min}$ for increasing $N$ (case $H_1$, $\alpha = 1/10$).}
 \label{fig_h1_asymp_0.1}
\end{figure}

Figures \ref{fig_h1_asymp_0.5} and \ref{fig_h1_asymp_0.1} show the
behavior of the $\lambda_{max}/\lambda_{min}$ for the signal plus
noise case for $\alpha = 1/2$ and $\alpha = 1/10$, respectively. In
both cases, $\sigma^2=1/\rho$ (with a $\rho$ of -5~dB) with
$\sum{\vert h_i\vert^2}=1$ (which holds under the criteria in Eq.
(\ref{eq_baik_crit})). In this case,
$\frac{\lambda_{max}}{\lambda_{min}} = \frac{b'}{a}$, for the pure
signal case. Interestingly, for $N=100$ ($K=50$ for $\alpha = 1/2$
and $K=10$ for $\alpha = 1/10$), it can be seen that the simulated
case is approximately 70\% percent and 83\% of the asymptotic limit
for $\alpha = 1/2$ and $\alpha = 1/10$, respectively. As expected,
the larger the ${\bf Y}$ matrix sizes, the closer one gets to the
asymptotic ratio. A good approximation was obtained  for  values of
$N$ as low as 100 samples.

\section{Results}\label{sec_results}

Simulations were carried out to establish the performance of the
random matrix theory detector scheme in comparison to the
cooperative energy detector scheme based on voting
\cite{art-ghasemi-2005, art-ghasemi-2007}. The framework for the
energy detector is exposed in section \ref{sec_prob_formulation},
with $h(k)$ modeled as a rayleigh multipath fading of variance
$1/K$. The variance is normalized to take into account the fact that
the energy does not increase without bound as the number of base
stations increases due to the path loss. A total of 10 secondary
base stations were simulated. For the voting scheme, the decision
rule is the following: one considers the overall spectrum occupancy
decision to be the one chosen by most of the secondary base
stations. The threshold $V_T$ is taken as $\sigma^2$ (for the known
noise variance case). For the random matrix theory based scheme, a
fixed total of ($K=10$) base stations were adopted.  Note that  the
algorithms can be optimized for the voting and random matrix theory
based rules by adopting decision margins \cite{art-cardoso-2008}.

\begin{figure}[h]
 \centering
 \psfrag{Number of samples}{\tiny{Number of samples}}
 \psfrag{Proportion of correct detections}{\tiny{Proportion of correct detections}}
 \includegraphics[width=0.6\columnwidth]{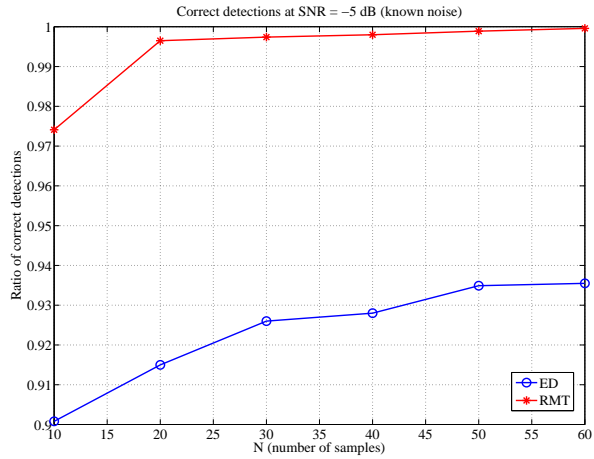}
 \caption{Comparison between the ED and random matrix theory approach  ($\rho=-5~dB$).}
 \label{fig_var_known}
\end{figure}

Figure \ref{fig_var_known} depicts the performance of the energy
detector scheme along with the random matrix theory one for $N=\{10,
20, ..., 60\}$ samples and a known noise variance of $\sigma^2$ at
SNR equal to -5dB. It is important to stress that since $K$ is
fixed, $\alpha$ is not constant as in the previous section. As
clearly shown, the random matrix theory scheme outperforms the
cooperative energy detector case for all number of samples due to
its inherent robustness.

\begin{figure}[h]
 \centering
 \psfrag{Number of samples}{\tiny{Number of samples}}
 \psfrag{Proportion of correct detections}{\tiny{Proportion of correct detections}}
 \includegraphics[width=0.6\columnwidth]{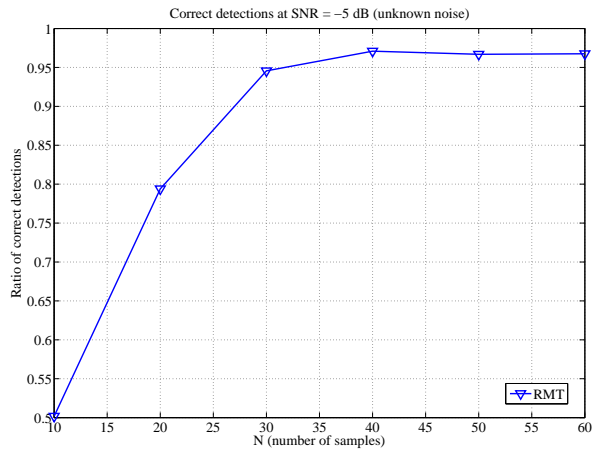}
 \caption{Random matrix theory approach for an unknown noise variance.}
 \label{fig_var_unknown}
\end{figure}

Figure \ref{fig_var_unknown} plots  the performance of the random
matrix theory scheme for an unknown noise variance (the voting
scheme can not be compared as it relies on the knowledge of the
noise variance). One can see that, indeed, even without the
knowledge of a noise variance, one is still able to achieve a very
good performance for sample sizes greater than 30.

\section{Conclusions}\label{sec_conclusions}

In this paper, we have provided a new spectrum sensing technique
based on random matrix theory and shown its performance in
comparison to the cooperative energy detector scheme for both a
known and unknown noise variance. Remarkably, the new technique is
quite robust and does not require the knowledge of the signal or
noise statistics. Moreover, the asymptotic claims turn out to be
valid even for a very low number of dimensions. The method can be
enhanced (see \cite{art-cardoso-2008}) by adjusting the threshold
decision, taking into account the number of samples though the
derivation of the probability of false alarm of the limiting  ratio
of the largest to the smallest eigenvalue.

\section*{Acknowledgment}
This work was funded by the Alcatel-Lucent Chair in Flexible Radio.
The authors would like to thank prof. Walid Hachem for  fruitful
discussions on the topic.



%
\bibliographystyle{unsrt}
\bibliography{rmt_ss}

\end{document}